\documentclass[aps,prl,preprint,superscriptaddress]{revtex4-1}

\usepackage{graphicx}
\usepackage{dcolumn}
\usepackage{bm}
\usepackage{epstopdf}

\begin{document}


\title{A Gapless MoS$_2$ Allotrope Possessing Both Massless Dirac and Heavy Fermions}


\author{Weifeng Li}
\affiliation{Institute of High Performance Computing, A*STAR, Singapore 138632.}

\author{Meng Guo} 
\affiliation{National Supercomputer Center in Jinan, Jinan, P. R. China 250101.}

\author{Gang Zhang}
\email[]{zhangg@ihpc.a-star.edu.sg}
\affiliation{Institute of High Performance Computing, A*STAR, Singapore 138632.}

\author{Yong-Wei Zhang}
\email[]{zhangyw@ihpc.a-star.edu.sg}
\affiliation{Institute of High Performance Computing, A*STAR, Singapore 138632.}


\date{\today}

\begin{abstract}
MoS$_2$, a member of transition metal dichalcogenides (TMDs), recently emerged as one of the fastest growing two-dimensional materials due to its fascinating mechanical, thermal, electronic and optical properties. Unlike graphene which possesses massless Dirac fermions with ultra-high electron mobility, monolayer MoS$_2$ is a direct band gap semiconductor. An interesting question arises: Can monolayer MoS$_2$ also possess massless Dirac fermions with ultra-high electron mobility? Here, using first-principles calculations, we show that a monolayer MoS$_2$ allotrope, which consists of repeated square-octagon rings (abbreviated as so-MoS$_2$ to distinguish from the normal hexagonal lattice, h-MoS$_2$) possesses both massless Dirac fermions and heavy fermions. Distinct from the $p$-orbital Dirac fermions of graphene, the Dirac fermions of so-MoS$_2$ are $d$-electrons and possess Fermi velocity comparable to that of graphene. The Dirac cone structure in so-MoS$_2$ demonstrated here greatly enriches our understanding on the physical properties of TMDs and opens up new possibilities for developing novel electronic/spintronic devices.
\end{abstract}



\maketitle

\section{Introduction}
One of the exceptional physical properties of graphene is the so-called Dirac cone structure\cite{Neto2009}. In the reciprocal space, the valence and conduction bands of graphene, which follow a linear dispersion relation confined to two dimensions, meet at a single point (called Dirac point) at the Fermi level (E$_{\rm{F}}$). This peculiar band structure results in an extremely high charge carrier mobility in the form of massless Dirac fermions, and also many other interesting phenomena, ranging from the formation of conducting edge states in topological insulators to the quantum spin Hall effect due to spin-orbit coupling\cite{Neto2009,Liu2011,Hasan2010,Qi2010,Zhang2009}. Since then, a great deal of effort has been made to search for new materials that also possess Dirac cone. Apart from graphene, Dirac cones were also reported in several graphene derivatives\cite{Malko2012,Gomes2012,Li2012,Hwang2012}, silicene and hexagonal germanium and their hybrids\cite{Liu2011,Zhou2013}. Common features of these 2D Dirac systems are: 1) a honeycomb lattice structure; and 2) $sp^2$ hybridized orbitals near E$_{\rm{F}}$ from group IV elements. It is commonly believed that the formation of Dirac cone arises from the strong interplay between electronic and structural properties. For all the known 2D Dirac systems, the three pear-shaped lobes of the $sp^2$ hybridized orbitals are always in steric compatibility with the honeycomb topology, with no exception.

In recent years, monolayer MoS$_2$, a member of transition metal dichalcogenides (TMDs), emerged as one of the fastest growing 2D materials\cite{Chhowalla2013}. Monolayer MoS$_2$ shares a similar planar hexagonal lattice to graphene, but its frontier orbitals are dominated by $d$-electrons. In contrast to graphene, monolayer MoS$_2$ sheet is a semiconductor with no Dirac cone. Although the allotropes of graphene have been studied extensively\cite{Malko2012,Gomes2012,Xu2014}, the allotropes of monolayer MoS$_2$ have not been explored. Then, a couple of interesting questions arises: Is there any stable allotrope of monolayer MoS$_2$? If yes, does it possess a Dirac cone electronic structure?

Inspired by recent theoretical and experimental studies of grain boundary atomic structures of monolayer MoS$_2$\cite{Najmaei2013,Zande2013,Zhang2013}, we first construct a monolayer MoS$_2$ allotrope with a square crystal lattice. We then reveal that this new monolayer material possesses a nontrivial Dirac cone with massless d electrons. Finally, we perform energetic and phonon spectrum analyses to prove that this novel material is structurally stable. To our best knowledge, this is the first demonstration of Dirac fermions in a monolayer crystal beyond the group IV elements and honeycomb lattice symmetry.

\section{Calculation methods}
All the calculations were performed using Vienna ab-initio simulation package (VASP)\cite{Kresse1996,Kresse1996_2}. Projector-augmented-wave (PAW) potentials\cite{Blochl1994} were used to take into account the electron-ion interactions, while the electron exchange-correlation interactions were treated using generalized gradient approximation (GGA)\cite{Perdew1992} in the scheme of Perdew-Burke-Ernzerhof. A plane wave cutoff of 500 eV was used for all the calculations. All atomic positions and lattice vectors were fully optimized using a conjugate gradient algorithm to obtain the ground-state configuration. A vacuum layer of 25 \AA~were kept between so-MoS$_2$ sheets in the \textbf{\textit{c}} direction to avoid mirror interactions. Atomic relaxation was performed at extremely critical conditions: the change of total energy was less than 0.001 meV and all the Hellmann-Feynman forces on each atom were smaller than 0.0001 eV/\AA, which guarantee fully relaxed structures and the accuracy for the following phonon dispersion calculation. A $k$-point sampling of $7\times7\times1$ was used for the structure relaxation, while a denser mesh of $45\times45\times1$ was used to calculate density of states (DOS) and band structures.

The phonon dispersion was calculated using the PHON program\cite{alfe2009phon} based on the finite displacement method. A super-cell containing $3\times3$ of the primitive-cell was used to calculate the force constant matrix. According to the symmetry operations of $p4$ crystal, there are two in-equivalent atoms in the primitive-cell (one Mo and one S as shown in Fig.~\ref{fig:struc}A and the displacements along the \textbf{\textit{a}}-axis and \textbf{\textit{c}}-axis were considered for each atom (with displacement of 0.04 \AA). The forces on all the atoms in the super-cell for each displacement were calculated with VASP with a $k$-point sampling of $9\times9\times1$.

\section{Results and Discussion}
\begin{figure}
\includegraphics[height=6cm]{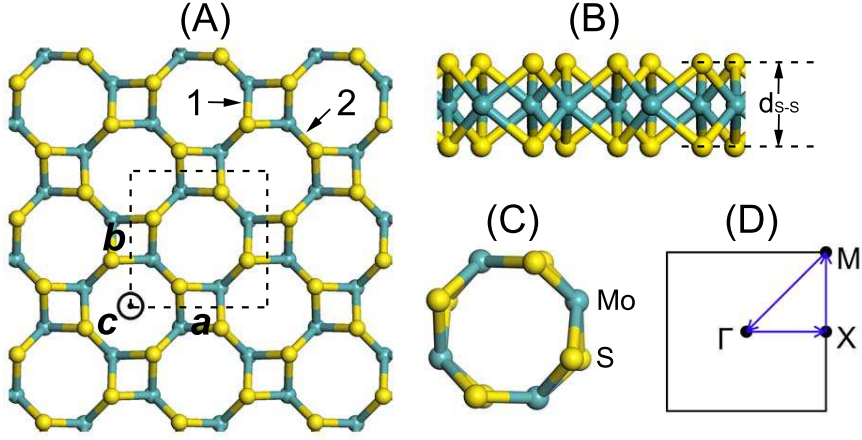}
\caption{\label{fig:struc} (A) Top view and (B) side view of so-MoS$_2$. The dash-line square and bold italic text \textbf{\textit{a}}, \textbf{\textit{b}} and \textbf{\textit{c}} in (A) indicate the primitive cell which is also highlighted in (C) and the three lattice vector directions. (D) The first Brillouin zone of so-MoS$_2$ with letters designating special points and the lines for band structure calculation.}
\end{figure}

In commonly studied monolayer MoS$_2$, that is, h-MoS$_2$, the transition metal atoms locate in one sublattice, which is sandwiched by two superimposed chalcogenide atomic layers in the other sublattice. In addition to 6-membered rings, 4- and 8-membered rings consisting entirely of hetero-elemental bonds (S-Mo) were also found at the grain boundaries in monolayer MoS$_2$\cite{Zande2013}. Inspired by these stable grain boundary structures, we construct a monolayer MoS$_2$ as shown in Fig.~\ref{fig:struc}A by repeating the square-octagon pairs in a square lattice (abbreviated as so-MoS$_2$ to distinguish from the normal hexagonal lattice, h-MoS$_2$). The primitive cell is square with a $p4$ symmetry, containing four Mo and eight S atoms.

The optimized structure of so-MoS$_2$ has a lattice constant of 6.36 \AA. The thickness of the monolayer (measured from S in the top layer to S in the bottom layer as illustrated in Fig.~\ref{fig:struc}B) is 3.12 \AA, similar to that of h-MoS$_2$ which is 3.13 \AA. There are structural distortions in so-MoS$_2$ because of the square-octagon topology: the Mo-S bonds that constitute the square (labelled 1 in Fig.~\ref{fig:struc}A) are stretched to 2.46 \AA, while the Mo-S bonds that link the squares (labelled 2 in Fig.~\ref{fig:struc}A) are compressed to a length of 2.39 \AA. As a comparison, the Mo-S bond in h-MoS$_2$ is homogeneously 2.42 \AA.

\begin{figure*}
\includegraphics[height=5cm]{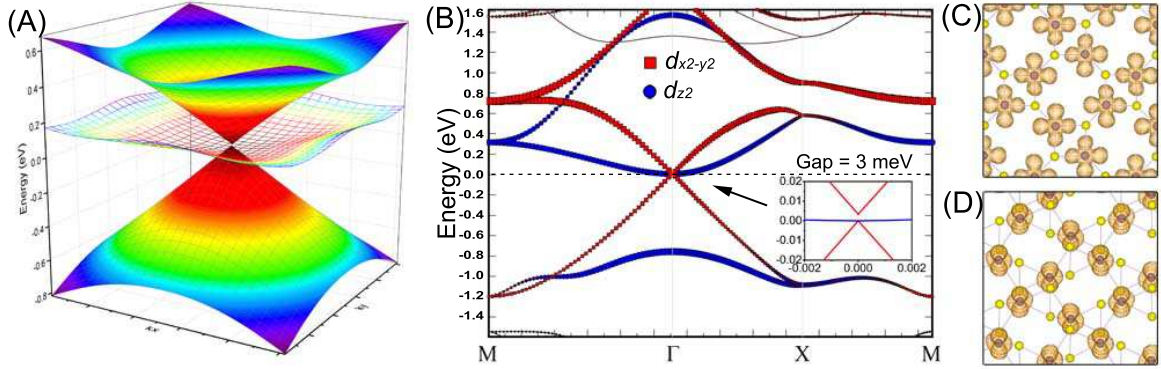}
\caption{\label{fig:elec} (A) Dirac cone formed in the vicinity of Dirac point. (B) Electronic band structures of so-MoS$_2$.  (C, D) Charge density plots of the $d_{x2-y2}$ and $d_{z2}$ orbitals of Mo, respectively (electron orbital isosurface level of 0.001 \AA$^{-3}$).}
\end{figure*}

A representative 3D plot that reveals the shape of the conduction and valence bands is shown in Fig.~\ref{fig:elec}A. It is exciting to find that the lower cone (from the valence band) possesses a linear dispersion relation near the Dirac point, indicating massless Dirac fermions. The electronic band structures of so-MoS$_2$ are shown in Fig.~\ref{fig:elec}B. It is seen that the valence band, conduction band and conduction band+1 (VB, CB and CB+1) approach the $\Gamma$ point at E$_{\rm{F}}$. The slopes of VB are -16.0 eV$\cdot$\AA (X to $\Gamma$) and +15.4 eV$\cdot$\AA (M to $\Gamma$), respectively. The upper cone (from CB+1) locates 3 meV above E$_{\rm{F}}$ at the $\Gamma$ point. The slopes of CB+1 branch are +15.6 eV$\cdot$\AA (X to $\Gamma$) and -15.5 eV$\cdot$\AA (M to $\Gamma$), respectively. We also calculate the Fermi velocity ($v_{\rm{F}}$) of the Dirac fermions using $v_{\rm{F}} = \frac{E(k)}{\hbar k}$, which is about 2.3-2.4$\times$10$^6$ m/s, comparable to that of graphene\cite{Hwang2012,Avouris2007}, but an order of magnitude higher than those of hexagonal silicene and germanium monolayer\cite{Liu2011,Zhou2013}. The CB deviates from the linear slope and is flattened, indicating the coexistence of heavy fermions. The effective mass of these fermions can be estimated using the following equation at the CB minimum.

\begin{eqnarray}
m^* = \frac{\hbar^2}{\partial^2 E(k)/\partial k^2}
\end{eqnarray}

\begin{figure}
\includegraphics[height=9cm]{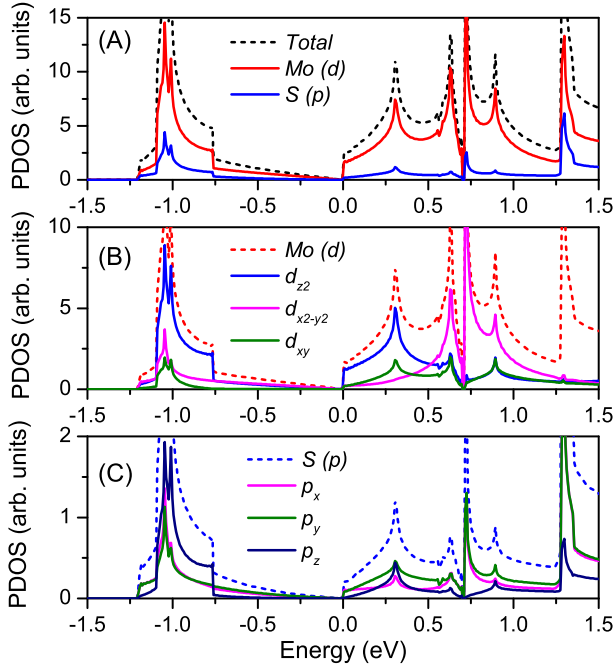}
\caption{\label{fig:dos} (A) Total DOS and projections on Mo $d$- and S $p$-orbitals; (B) PDOS of Mo $d$-orbitals and projections on $d_{z2}$, $d_{x2-y2}$ and $d_{xy}$; (C) PDOS of S $p$-orbitals and projections on $p_x$, $p_y$ and $p_z$.}
\end{figure}

Our calculations show that the effective masses are 1.11 $m_e$ (X to $\Gamma$) and 1.78 $m_e$ (M to $\Gamma$), respectively. The complicity of these two un-occupied bands contributes to a moderate level of density of state (DOS) just above E$_{\rm{F}}$, while the DOS just below E$_{\rm{F}}$ is zero as shown in Fig.~\ref{fig:dos}A. This implies that, above zero temperature, excitations of electrons to CB and CB+1 are able to cause a high charge carrier density as well as extremely high carrier mobility. Hence a better conductivity than graphene is expected due to its zero carrier density at E$_{\rm{F}}$.

It is well-known that in TMDs, the frontiers orbitals are mainly originated from the $d$-orbitals of transition metals, with weak contribution from the $p$-orbitals of chalcogen atoms\cite{Kuc2011,Ding2011,voss1999atomic,Albe2002,Han2011}. In h-MoS$_2$, $d_{xy}$ and $d_{x2-y2}$ orbitals of Mo are doubly-degenerated in energy, while $d_{yz}$ and $d_{xz}$ are also doubly-degenerated and $d_{z2}$ is singly-denegerated. For S atoms, the $p_x$ and $p_y$ are degenerated. From hexagonal lattice ($p6m$ summatry) to square lattice ($p4$ summatry), some structure symmetries are lost, as well as the electron orbital degeneracies. These changes can be clearly observed from Fig.~\ref{fig:dos}. In so-MoS$_2$, the VB and CB+1 are mainly described by the $d_{x2-y2}$ orbital of Mo. In the real space, the four pear-shaped lobes of $d_{x2-y2}$ orbital spread in the so-MoS$_2$ plane and have centres on the \textbf{\textit{a}} and \textbf{\textit{b}} lattice axes as shown in Fig.~\ref{fig:elec}C, fully compatible with the square lattice symmetry. Consequently, a long-range coherence is realized between the $d_{x2-y2}$ orbital and the square crystal lattice, leading to the formation of Dirac cone. On the other hand, the CB branch is mainly described by the out-of-plane $d_{z2}$ orbital of Mo. The two pear-shaped regions of $d_{z2}$ are more localized, placing symmetrically on the \textbf{\textit{c}} axis as shown in Fig.~\ref{fig:elec}D, resulting in the heavy fermions.

\begin{figure}
\includegraphics[height=7cm]{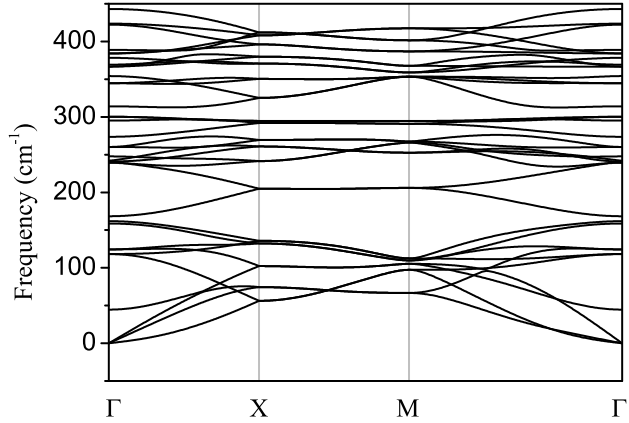}
\caption{\label{fig:phon} Phonon band structures of so-MoS$_2$ along high symmetry directions in the Brillouin zone.}
\end{figure}

In order to assess the stability of the so-MoS$_2$-based Dirac Fermion system, we first calculated the formation energy (normalized to each atom) defined as:

\begin{eqnarray}
E_{form} = \frac{E_{total} - n_{Mo} \times E_{Mo} - n_{S} \times E_{S}}{n_{Mo} + n_{S}}
\end{eqnarray}

Our calculation shows that $E_{form}$ is about -4.79 eV/atom, comparable to a value of -5.07 eV/atom for h-MoS$_2$, indicating that so-MoS$_2$ is only slightly less stable than h-MoS$_2$. Furthermore, we also examined the thermal stability by calculating the phonon dispersion relations, which can pinpoint the stability of materials\cite{Peelaers2008}. The phonon band structures are shown in Fig.~\ref{fig:phon}. It is seen that originating from the $\Gamma$ point, there are three zero-frequency acoustic branches. The lowest branch shows a quadratic energy dispersion relation in $k$, corresponding to the out-of-plane displacement. The other two branches display a linear $k$-dependence around the $\Gamma$ point, representing the in-plane translational and longitudinal displacements in two orthogonal directions. As there are no imaginary frequencies (negative frequencies) found in the phonon spectrum, hence the so-MoS$_2$ is structurally stable. The possible approach to realize so-MoS$_2$ is to use novel substrates, like growing hexagonal hafnium (Hf) layer on Ir(111) surface\cite{Li2013} and silicene on Ag(111)\cite{Feng2012,Vogt2012}.

\section{Conclusion}
To conclude, for the first time, massless Dirac fermions are observed in a member of TMD family, beyond the group IV elements and honeycomb lattice symmetry. We find that the $d_{x2-y2}$ orbitals in so-MoS$_2$ form a long-range coherence with square lattice structure, leading to the formation of Dirac fermions. We also find that the $d_{z2}$ orbitals locate near E$_{\rm{F}}$, resulting in the coexistence of heavy fermions with massless Dirac fermions in so-MoS$_2$. Our findings here undoubtedly will not only attract great interest in search for new Dirac fermion materials and their associated new physical properties but also create a world of new possibilities for developing novel electronic, photonic and spintronic atomically-thin layered devices through lattice engineering.

\section{Supplemental Material}
Phonon band structures of so-MoS$_2$ calculated with the linear response method, and change density plots of the four orbitals near the Fermi level can be found in the Supplemental Material. This material is available at http://XXXXXXX.

\begin{acknowledgments}
This work was supported by the A*STAR Computational Resource Centre through the use of its high performance computing facilities.
\end{acknowledgments}

\bibliography{mos2_dirac}

\end{document}